\documentclass{article}
\usepackage{spconf,amsmath,graphicx,amssymb,amsthm,comment,bm,cite,url}
\usepackage{array}
\allowdisplaybreaks

\newcommand{\refeq}[1]{(\ref{eq:#1})}
\newcommand{\refeqs}[2]{(\ref{eq:#1}) and (\ref{eq:#2})}

\newcommand{\reffig}[1]{Fig. \ref{fig:#1}}

\def\Vec#1{\boldsymbol{\mathbf{#1}}}

\def\Argmin{\mathop{\rm argmin}}

\makeatletter
\newcommand{\thickhline}{%
    \noalign {\ifnum 0=`}\fi \hrule height 1pt
    \futurelet \reserved@a \@xhline
}
\newcolumntype{"}{@{\hskip\tabcolsep\vrule width 1pt\hskip\tabcolsep}}
\makeatother

\title{
Generative adversarial network-based approach to signal reconstruction from magnitude spectrograms
}
\name{
\begin{tabular}{c}
Keisuke Oyamada$^\star$, Hirokazu Kameoka$^\dagger$, Takuhiro Kaneko$^\dagger$,\\
Kou Tanaka$^\dagger$, Nobukatsu Hojo$^\dagger$, Hiroyasu Ando$^\star$
\end{tabular}
}
\address{
$^\star$University of Tsukuba, Japan\\
$^\dagger$NTT Communication Science Laboratories, NTT Corporation, Japan
}
\begin{document}
\maketitle
\begin{abstract}
In this paper, we address the problem of reconstructing a time-domain signal (or a phase spectrogram) solely from a magnitude spectrogram.
Since magnitude spectrograms do not contain phase information, we must 
restore or infer phase information to reconstruct a time-domain signal.
One widely used approach for dealing with the signal reconstruction problem was proposed by Griffin and Lim.
This method usually requires many iterations for the signal reconstruction process and 
depending on the inputs, it does not always produce high-quality audio signals.
To overcome these shortcomings, we apply a learning-based approach to the signal reconstruction problem by modeling the signal reconstruction process using a deep neural network and training it using the idea of a generative adversarial network.
Experimental evaluations revealed that our method was able to reconstruct signals 
faster with higher quality than the Griffin-Lim method.
\end{abstract}
\section{Introduction}
This paper addresses the problem of reconstructing a time-domain signal solely from a magnitude spectrogram.

The magnitude spectrograms of real-world audio signals tend to be highly structured
in terms of both spectral and temporal regularities. 
For example, pitch contours and formant trajectories are clearly visible from a magnitude spectrogram representation of speech compared with a time-domain signal.  
Therefore, there are many cases where processing magnitude spectrograms can 
deal with problems more easily than directly processing time-domain signals.
In fact, many methods for monaural audio source separation are applied to
magnitude spectrograms \cite{Smaragdis2014,Virtanen2015,Kameoka2016}.
Furthermore, a magnitude spectrogram representation 
was recently found to be reasonable and effective for use with speech synthesis systems \cite{Takaki2017,Wang2017}.

Since a magnitude spectrogram does not contain phase information, 
we must restore or infer phase information to 
reconstruct a time-domain signal. This problem is called the signal (or phase) reconstruction problem. 
One widely used method for solving the signal reconstruction problem 
was proposed by Griffin and Lim \cite{Griffin1984} (hereafter referred to as the Griffin-Lim method).
One of the drawbacks of the Griffin-Lim method is that it usually requires many iterations to obtain high-quality audio signals. This makes it particularly difficult to apply it to  real-time systems.
Furthermore, there are some cases where
high-quality audio signals can never be obtained
even though the algorithm is run for many iterations.
To overcome these shortcomings of the Griffin-Lim method, 
we apply a learning-based approach to the signal reconstruction problem.
Specifically, 
we propose modeling the reconstruction process of a time-domain signal from a magnitude spectrogram using a deep neural network (DNN) and
propose introducing the idea of the generative adversarial network (GAN) \cite{Goodfellow2014} for training the signal generator network.

The remainder of the paper is organized as follows.
We provide an overview of the phase reconstruction problem in Section \ref{prp},  
introduce the Griffin-Lim method in Section \ref{glm}, and
present our GAN-based approach in Section \ref{pm}.
Experimental evaluations, and supplements for training our model are provided in Section \ref{ee}.
Finally, we offer our conclusions in Section \ref{conc}.

\section{Signal Reconstruction Problem}
\label{prp}
In this section, we provide an overview of the signal reconstruction problem.

We use 
 $\Vec{x}=[x(0),\ldots,x(T-1)]^{\mathsf T}\in\mathbb{R}^{T}$ to denote 
a time domain signal and $c_{f,n}\in\mathbb{C}$ to denote
the time frequency component of $\Vec{x}$ 
where $f$ and $n$ indicate frequency and time indices, respectively.
By defining
$\Vec{w}_{f,n}=[w_{f,n}(0),\ldots,w_{f,n}(T-1)]^{\mathsf T}\in\mathbb{C}^{T}$
as a complex sinusoid of frequency $\omega_f$ modulated by a window function centered at time $t_n$, 
$c_{f,n}$ is defined by the inner product between 
$\Vec{x}$ and $\Vec{w}_{f,n}$, namely $c_{f,n} = \Vec{w}_{f,n}^{\mathsf H}\Vec{x}$. 
With a short-time Fourier transform (STFT), $t_n$ corresponds to the center time of frame $n$ and
$\Vec{w}_{f,n}$ is the modulated complex sinusoid padded with zeros over the range outside the frame. 
By using $\Vec{c}\in\mathbb{C}^{FN}$ to denote a vector 
obtained by stacking all the time-frequency components $c_{f,n}$, 
the relationship between 
$\Vec{c}$ and $\Vec{x}$ can be written as
\begin{align}
\Vec{c} = \Vec{W}\Vec{x},
\label{eq:TFredundancy}
\end{align}
where $\Vec{W}$ is a $FN \times T$ matrix where each row is $\Vec{w}_{f,n}^{\mathsf H}$.
Hereafter, we call $\Vec{c}$ a complex spectrogram.
Since the total number $FN$ of time frequency points is usually set at more than the number $T$ of sample points of the time domain signal, 
$\Vec{c}$ is a redundant representation of $\Vec{x}$.
Namely, $\Vec{c}$ belongs to a $T$-dimensional linear subspace $\mathcal{C}$ spanned by each column vector of $\Vec{W}$.
With an STFT, 
all the elements of
a complex spectrogram must satisfy certain conditions to
ensure that the waveforms within the overlapping segment of
consecutive frames are consistent.
By using $\Vec{a}$ to denote the magnitude spectrogram of $\Vec{c}$ 
where each element of $\Vec{a}$ is 
given by the absolute value of the element of $\Vec{c}$, 
the signal reconstruction problem can be cast as an optimization problem
of estimating $\Vec{x}$ solely from $\Vec{a}$ using the redundancy constraint 
as a clue.

\section{Griffin-Lim Method}
\label{glm}
One widely used way of solving the phase reconstruction problem
involves the Griffin-Lim method \cite{Griffin1984}.
In this section, we derive the iterative algorithm of the Griffin-Lim method 
following the derivation given in \cite{LeRoux2010}.

Whether or not a given 
$\Vec{c}$ satisfies the redundancy constraint so that
$\Vec{c}$ is a complex spectrogram associated with a time domain signal 
can be evaluated by examining whether or not the orthogonal projection $\Vec{W}\Vec{W}^{+}\Vec{c}$ of $\Vec{c}$ to the subspace $\mathcal{C}$ matches $\Vec{c}$.
Here, $\Vec{W}^{+}$ is a pseudo inverse matrix of $\Vec{W}$ satisfying 
\begin{align}
\Vec{W}^{+}\Vec{c} &= \Argmin_{\Vec{x}} \|\Vec{c} - \Vec{W}\Vec{x}\|_2^2
\nonumber\\
&
= (\Vec{W}^{\mathsf H}\Vec{W})^{-1}\Vec{W}^{\mathsf H}\Vec{c}.
\label{eq:pseudoinverseW}
\end{align}
With an STFT,
\refeq{pseudoinverseW} 
corresponds to an inverse STFT.
Thus, $\Vec{W}\Vec{W}^{+}\Vec{c}$ is the STFT of
the inverse STFT of $\Vec{c}$. 
Now, by using $\Vec{\phi}$ to denote a vector where each element 
is the phase $\phi_{f,n}\equiv e^{\j\theta_{f,n}}$, 
the phase reconstruction problem for a given $\Vec{a}$ is formulated 
as an optimization problem of estimating $\Vec{\phi}$ 
that minimizes 
\begin{align}
\mathcal{J}(\Vec{\phi}) = 
\|\Vec{\Vec{a}\odot \Vec{\phi}} - \Vec{W}\Vec{W}^{+}(\Vec{\Vec{a}\odot \Vec{\phi}})\|_2^2,
\label{eq:optimizationPhai}
\end{align}
where $\odot$ denotes an element-wise product.
Now, from \refeq{pseudoinverseW}, $\Vec{W}\Vec{W}^{+}(\Vec{\Vec{a}\odot \Vec{\phi}})$ is 
the point closest to $\Vec{\Vec{a}\odot \Vec{\phi}}$ in the subspace $\mathcal{C}$.
Thus, we can rewrite \refeq{optimizationPhai} as
\begin{align}
\mathcal{J}(\Vec{\phi}) &= 
\min_{\tilde{\Vec{c}}\in\mathcal{C}} 
\|\Vec{\Vec{a}\odot \Vec{\phi}} - \tilde{\Vec{c}}\|_2^2.
\label{eq:auxiliaryfunc}
\end{align}
According to the principle of the majorization-minimization algorithm \cite{Hunter2000}, 
it can be shown that
$\mathcal{J}^+(\Vec{\phi},\tilde{\Vec{c}})\equiv \|\Vec{\Vec{a}\odot \Vec{\phi}} - \tilde{\Vec{c}}\|_2^2$ is a majorizer of $\mathcal{J}(\Vec{\phi})$ 
where $\tilde{\Vec{c}}\in\mathcal{C}$ is an auxiliary variable
and a stationary point of $\mathcal{J}(\Vec{\phi})$ can be found 
by iteratively performing the following updates:
\begin{align}
\tilde{\Vec{c}} &\leftarrow \Argmin_{\tilde{\Vec{c}}\in\mathcal{C}} 
\|\Vec{\Vec{a}\odot \Vec{\phi}} - \tilde{\Vec{c}}\|_2^2 =
\Vec{W}\Vec{W}^{+}(\Vec{\Vec{a}\odot \Vec{\phi}}),
\label{eq:GLalgo_step1}
\\
\Vec{\phi}&\leftarrow 
\Argmin_{\Vec{\phi}} 
\|\Vec{\Vec{a}\odot \Vec{\phi}} - \tilde{\Vec{c}}\|_2^2
=\angle \tilde{\Vec{c}}.
\label{eq:GLalgo_step2}
\end{align}
Here $\angle \cdot$ denotes an operation that divides each element of a vector by its absolute value.
With an STFT, 
Eq. \refeq{GLalgo_step1} can be interpreted as the inverse STFT of $\Vec{\Vec{a}\odot \Vec{\phi}}$ followed by the STFT whereas
Eq. \refeq{GLalgo_step2} is a procedure for replacing the phase $\Vec{\phi}$
with the phase of $\tilde{\Vec{c}}$ updated via \refeq{GLalgo_step1}.
This algorithm is procedurally equivalent to the Griffin-Lim method \cite{Griffin1984}.

The Griffin-Lim method usually requires many iterations 
to obtain a high-quality audio signal. This makes 
it particularly difficult to apply to real-time systems.
Furthermore, there are some cases where
high-quality audio signals can never be obtained
even though the algorithm is run for many iterations, for example when $\Vec{a}$ is an artificially created magnitude spectrogram.
In the next section, we propose a learning-based approach to the
phase reconstruction problem to overcome these shortcomings of the Griffin-Lim method.

\section{GAN-based signal reconstruction}
\label{pm}
\subsection{Modeling phase Reconstruction Process}
By using $\Vec{\phi}^{(0)}$ to denote 
the initial value of $\Vec{\phi}$, and defining
$h(\Vec{a},\Vec{\phi}) \equiv \Vec{W}\Vec{W}^{+}\Vec{a}\odot\Vec{\phi}$ and
$g(\Vec{c})\equiv \angle \Vec{c}$, 
the iterative algorithm of the Griffin-Lim method can be expressed as a multilayer composite function
\begin{align}
\hat{\Vec{c}} = h(\Vec{a},g(\cdots g(h(\Vec{a},g(h(\Vec{a},\Vec{\phi}^{(0)}))))\cdots)).
\label{eq:DNNinterpretation}
\end{align}
Here, 
$h$ is a linear projection whereas $g$ is a nonlinear operation applied to the output of $h$.
Hence,
\refeq{DNNinterpretation} can be viewed as a deep neural network (DNN) 
where the weight parameters and the activation functions are fixed.
From this point of view, finding an algorithm that converges more quickly to a better solution than the Griffin-Lim algorithm can be regarded as learning the
weight parameters (and the activation functions) of the DNN. 
This idea is inspired by the deep unfolding framework \cite{Hershey2014}, 
which uses a learning strategy to obtain an improved version of a deterministic
iterative inference algorithm by unfolding the iterations and treating them as layers in a DNN.  
Fortunately, an unlimited number of 
pair data of $\Vec{c}$ and $\{\Vec{a}, \Vec{\phi}\}$ 
can be collected very easily by computing 
the complex, magnitude and phase spectrograms of time domain signals.
This is very advantageous for efficiently training our DNN.

In the following, we consider a DNN that uses $\Vec{a}$ and $\Vec{\phi}$ as inputs and generates $\Vec{c}$ (or $\Vec{x}$) as an output. 
We call this DNN a generator $G$ and express the relationship between 
the input and output as
$\hat{\Vec{c}} = G(\Vec{a},\Vec{\phi})$.

\subsection{Learning Criterion}

For the generator training,
one natural choice for the learning criterion
would be a similarity metric (e.g., the $\ell_1$ norm) between 
the generator output and a target complex spectrogram (or signal). 
Manually defining a similarity metric 
amounts to assuming a specific form of the probability distribution 
of the target data (e.g., a Laplacian distribution for the $\ell_1$ norm).
However, the data distribution is unknown.
If we use a similarity metric defined in the data space as the learning criterion, 
the generator will be trained in such a way that the outputs that averagely fit  
the target data are considered optimal.
As a result, the generator will learn to generate only oversmoothed signals. 
This is undesirable as the oversmoothing of reconstructed signals causes audio quality degradation.
To avoid this, we propose using a similarity metric implicitly 
learned using a generative adversarial network (GAN) \cite{Goodfellow2014}.
In addition to the generator network, we introduce a discriminator network $D$ 
that learns to correctly discriminate the complex spectrograms $\hat{\Vec{c}}$ generated by the generator and the complex spectrograms of real audio signals. 
Given a target complex spectrogram $\Vec{c}$, 
the discriminator $D$ is expected to find a feature space where 
$\hat{\Vec{c}}$ and $\Vec{c}$ are as separate as possible. 
Thus, we expect that 
minimizing the distance between $\hat{\Vec{c}}$ and $\Vec{c}$ 
measured in a hidden layer of the discriminator 
would make $\hat{\Vec{c}}$ indistinguishable from $\Vec{c}$
in the data space. 
By using $D(\cdot,\Vec{a})\in\mathbb{R}$ to denote the discriminator network $D$,
we first consider the following criteria for the discriminator
\begin{align}
V(D) =& 
\frac{1}{2} \mathbb{E}_{(\Vec{c}, \Vec{a}) \sim p_{\Vec{c}, \Vec{a}}(\Vec{c}, \Vec{a})} 
\big[(D(\Vec{c},\Vec{a}) - 1)^2\big] \nonumber\\
+&
\frac{1}{2} \mathbb{E}_{\Vec{a} \sim p_{\Vec{a}}(\Vec{a}), \Vec{\phi} 
\sim p_{\Vec{\phi}}(\Vec{\phi})}
\big[D(G(\Vec{a}, \Vec{\phi}), \Vec{a})^2\big]. 
\label{eq:LSGAN1}
\end{align}
Here, the target label corresponding to real data is assumed to be 1 
and that corresponding to the data generated by the generator $G$ is 0. 
Thus, \refeq{LSGAN1} means that $V(D)$ becomes 0 only 
if the discriminator $D$ correctly distinguishes the ``fake'' complex spectrograms 
generated by the generator $G$ and the ``real'' complex spectrograms of real audio signals. 
Therefore, the goal of $D$ is to minimize $V(D)$.
As for the generator $G$, 
one of the goals is to deceive the discriminator $D$ so as to
make the ``fake'' complex spectrograms 
as indistinguishable as possible from the ``real'' complex spectrograms. 
This can be accomplished by minimizing the following criterion
\begin{align}
U(G) =&
\frac{1}{2} \mathbb{E}_{\Vec{a} \sim p_{\Vec{a}}(\Vec{a}), \Vec{\phi} 
\sim p_{\Vec{\phi}}(\Vec{\phi})}
\big[(D(G(\Vec{a}, \Vec{\phi}), \Vec{a})-1)^2\big].\!\!
\label{eq:LSGAN2}
\end{align}
Another goal for $G$ is to make $\hat{\Vec{c}}=G(\Vec{a},\Vec{\phi})$ as close as possible to the target complex spectrogram $\Vec{c}$.
By using $D_l(\cdot)$ to denote the output of the $l$-th layer of the discriminator $D$, 
we would also like $G$ to minimize 
\begin{align}
I(G) = \sum_{l=0}^{L}w_l \| D_l(\Vec{c}) - D_l(G(\Vec{a}, \Vec{\phi}))\|_2^2,
\label{eq:LSGAN_G}
\end{align}
where $w_l$ is a fixed weight, which weighs the importance of the $l$-th layer feature space. 
Here, the $0$-th layer corresponds to the input layer, namely $D_0(\Vec{c}) = \Vec{c}$.

The learning objectives for $D$ and $G$ 
can thus be summarized as follows:
\begin{align}
D:&~V(D) \rightarrow {\rm minimize},
\\
G:&~U(G)+\lambda I(G) \rightarrow {\rm minimize},
\label{eq:LSGAN_G_last}
\end{align}
where $\lambda$ is a fixed weight.

A general framework for training
a generator network in such a way that it can 
deceive a real/fake discriminator network 
is called a generative adversarial network (GAN) \cite{Goodfellow2014}.
The novelty of our proposed approach is that we have successfully  
adapted the GAN framework to
the signal reconstruction problem 
by incorporating an additional term \refeq{LSGAN_G}. 
The GAN framework using \refeqs{LSGAN1}{LSGAN2} as the learning criteria 
is called the least squares GAN (LSGAN) \cite{Mao2017}.
Note that GAN frameworks using 
other learning criteria such as \cite{Arjovsky2017} have also been proposed.
Thus, we can also use the learning criteria employed in \cite{Goodfellow2014}, \cite{Arjovsky2017} or others instead of \refeqs{LSGAN1}{LSGAN2}.

\section{Experimental Evaluation}
\label{ee}

We tested our method and the Griffin-Lim method using real speech samples.

\subsection{Experimental Settings}
\subsubsection{Dataset}
\label{test_config}
We used clean speech signals excerpted from \cite{valentini2016superseded} 
as the experimental data.
The speech data consisted of utterances of 30 speakers. 
The utterances of 28 speakers were used as the training set
and the remaining utterances were used as the evaluation set.
For the mini-batch training, we divided each training utterance into 1-second-long segments 
with an overlap of 0.5 seconds. 
All the speech data were downsampled to 16 kHz.
Magnitude spectrograms were obtained with an STFT
using a Blackman window that was 64 ms long with a 32 ms overlap.

\subsubsection{Network Architecture}
\begin{figure*}[tb]
\begin{center}
\includegraphics[width=2.0\columnwidth]{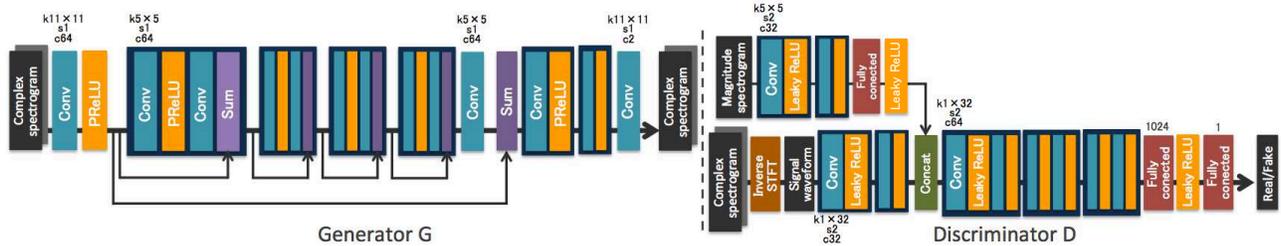}
\vspace{-17pt}     
\end{center}
\caption{
Network architectures of generator and discriminator. 
Light blue blocks indicate convolutional layers. 
In each convolutional layer, $\rm{k}$, $\rm{s}$, and $\rm{c}$ represent kernel size, stride size, and number of channels, respectively.
Here, $\rm{k} 1\times*$ indicates a one-dimensional convolutional layer whose kernel size is $*$.
Red blocks indicate fully connected layer.
In each fully connected layer, the numbers represents size of output unit.
}
\label{fig:model}
\vspace{-17pt}   
\end{figure*}

\reffig{model} shows the network architectures we constructed for this experiment.
The left half shows the architecture of the generator $G$ and the right half shows that of the discriminator $D$.
The light blue blocks indicate convolutional layers, and $\rm{k}$, $\rm{s}$, and $\rm{c}$ on each convolutional layer represent hyper-parameters.
The yellow blocks indicate activation functions.
PReLU\cite{he2015delving} was used for the generator $G$ and 
Leaky ReLU\cite{maas2013rectifier} was used for the discriminator $D$.
The violet blocks indicate element-wise sums, and 
the green block indicates the concatenation of features along the channel axis.
The red blocks indicate fully-connected layers.
Blocks without symbols have the same hyper-parameters as the previous blocks.
Note that we referred to \cite{ledig2016photo} when constructing these architectures.
The generator $G$ is fully convolutional \cite{long2015}, thus allowing an input to have an arbitrary length.
The weight constant $w_l$ was set to $0$ for $l = 0$ and $1$ for $l \neq 0$.
$\lambda$ was set to $1$.
RMSprop\cite{tieleman2012lecture} was used as the optimization algorithm and the learning rate was $5 \times 10^{-5}$C$\alpha=0.5$.
The mini-batch size was $10$ and the number of epochs was $73$.

Instead of directly feeding
an input magnitude spectrogram and a randomly-generated
phase spectrogram into the generator $G$, 
we used a complex spectrogram reconstructed 
using the Griffin-Lim method after 5 iterations as the $G$ input.
Both the input and output of the generator $G$ have 2 channels,
one corresponding to the real part 
and the other corresponding to the imaginary part 
of the complex spectrogram.
For pre-processing, we normalized the complex spectrograms of the training data
to obtain zero-mean and unit-variance at each frequency.
At test time, the scale of the generator output at each frequency was restored.

We added a block that applies an inverse STFT to the generator output before feeding it into the discriminator $D$.
We found this particularly important as 
the training did not work well without this block.

\subsection{Data Augmentation}

It is a well-known fact that 
the difference between signals 
is hardly perceptible to human ears
when the magnitude spectrograms and 
the inter-frame phase differences are the same. 
This implies that 
there is an arbitrariness in the initial phases of spectrograms 
that are perceived similarly. 
By utilizing this property, we can augment the training data for $G$ and $D$ by preparing 
many different waveforms that are the same except for the initial phases.
We expect that this data augmentation would allow the generator to concentrate on
learning a way of inferring  
appropriate inter-frame phase differences given a magnitude spectrogram, thus facilitating efficient learning. 

\subsection{Dimensionality Reduction}

Note that the real and imaginary parts of the Fourier transform of a real-valued signal 
become even and odd functions, respectively. 
Owing to this symmetric structure, 
it is sufficient to restore/infer spectral components within the frequency range from 0 up to the Nyquist frequency. We can therefore restrict the sizes of the input and output of the generator to this frequency range.

\subsection{Subjective Evaluation}
\label{sb_eval}
We compared our proposed method with 
the Griffin-Lim method in terms of the perceptual quality of 
reconstructed signals by conducting an AB test, 
where ``A'' and ``B'' were reconstructed signals obtained respectively with the
proposed and baseline methods. With this listening test, 
``A'' and ``B'' were presented in random orders to eliminate bias as regards the
order of stimuli. Five listeners participated in our listening
test. Each listener was presented with \{``A'',``B''\} 
$\times 10$ signals 
and asked to select ``A''or ``B'' for each pair.
The Griffin-Lim method was run for 400 iterations.
The signals were 2 to 5 seconds long. 

\begin{figure}[tb]
\begin{center}
\includegraphics[width=0.9\columnwidth]{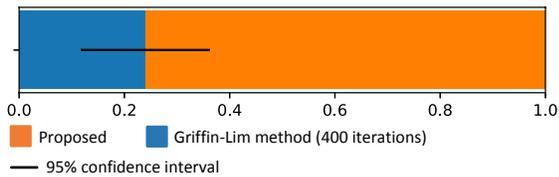}
\vspace{-17pt}     
\end{center}
\caption{
Result of the AB test. 
The orange area indicates the rate of the A and B pairs 
for which the listeners preferred A (proposed). 
The black bar indicates the 95\% confidence interval.
}
\label{fig:result_AB}
\vspace{-17pt}   
\end{figure}

The preference scores are shown in \reffig{result_AB}. 
As the result shows, the reconstructed signals obtained with 
the proposed method were preferred by the listeners 
for 76\% of the 50 pairs.

\subsection{Generalization ability}
\begin{figure}[tb]
\begin{center}
\includegraphics[width=0.9\columnwidth]{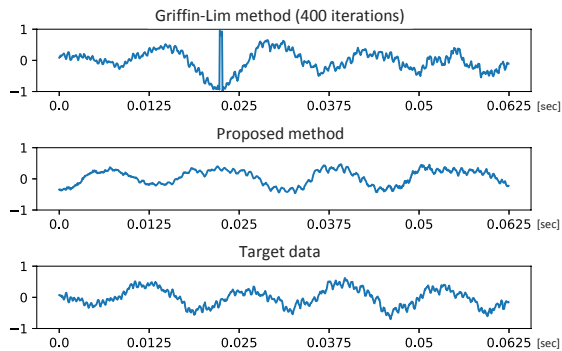}
\vspace{-17pt}     
\end{center}
\caption{
Waveforms of reconstructed music data \cite{jamendo}. 
The first row shows the acoustic signal reconstructed with the Griffin-Lim method, 
the second shows the proposed method, 
and the third is the target acoustic signal (real-world acoustic signal).
}
\label{fig:result_music}
\vspace{-17pt}   
\end{figure}

To confirm the generalization ability of the proposed method, 
we tested it on musical audio signals excerpted from \cite{jamendo}. 
Examples of the reconstructed signals are shown in \reffig{result_music}.
With these examples, we can observe a discontinuous point in
the reconstructed signal obtained with the Griffin-Lim method.
On the other hand, the proposed method 
appears to have worked successfully, even though the model was trained using speech data.

\subsection{Comparison of Processing Times}

\begin{figure}[tb]
\begin{center}
\includegraphics[width=0.9\columnwidth]{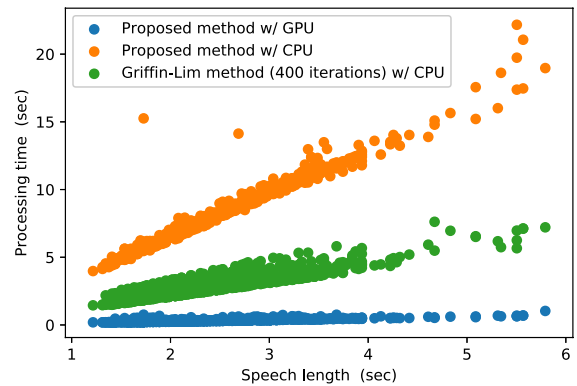}
\vspace{-17pt} 
\end{center}
\caption{
The change in processing time with respect to the change in speech length. 
Blue points show the processing time with the proposed method a GPU. 
Green points show the time with the Griffin-Lim method with a CPU.
Orange points show the time with the proposed method with a CPU.
}
\label{fig:result_time_all}
\vspace{-17pt}
\end{figure}

We further compared the proposed method with the Griffin-Lim method 
in terms of the processing times needed to reconstruct time domain signals.
For comparison, we measured the processing times for various speech lengths.
We used speech data shorter than 6 seconds for the evaluation.
Here, the network architecture of our proposed method was the same as \reffig{model}, 
and the Griffin-Lim method was run for 400 iterations.
The CPU used in this experiment was ``Intel Core i7-6850K CPU @ 3.60GHz''.
The GPU was ``NVIDIA GeForce GTX 1080''.
We implemented the Griffin-Lim method using the fast Fourier transform function in NumPy \cite{numpy}.
We implemented our model with Chainer \cite{chainer}.
\reffig{result_time_all} shows the result.
As the speech data become longer, the processing time increases linearly.
When executing the proposed method using the GPU, 
the time needed to reconstruct a signal was only about one-tenth the length of that signal.
On the other hand, the Griffin-Lim method executed using the CPU 
took about the same time as the length of the signal.
Therefore, if we can use a GPU, 
the proposed method can be run in real time. 
However, when using the CPU, the proposed method took about three times longer than the length of the signal. 
If we want to execute the proposed method in real-time using a CPU, we would need to 
construct a more compact architecture than that shown in \reffig{model}.
One simple way would be to replace the convolutional layers 
with downsampling and upsampling layers.

\section{Conclusion}
\label{conc}
This paper proposed a GAN-based approach to signal reconstruction from magnitude spectrograms. 
The idea was to model the signal reconstruction process using a DNN
and train it using a similarity metric implicitly learned using a GAN discriminator.
Through subjective evaluations,
we showed that the proposed method was able to 
reconstruct higher quality time domain signals
than the Griffin-Lim method, which was run for 400 iterations.
Furthermore, we showed that the proposed method can be executed in real-time when using a GPU.
Future work will include the investigation of a network architecture appropriate for CPU implementations.

\end{document}